\documentclass[prd,superscriptaddress,amsfonts,amssymb,amsmath,twocolumn,floatfix]{revtex4-2}
\usepackage{bm}
\usepackage{amsfonts}
\usepackage{latexsym}
\usepackage{graphicx}
\usepackage{amsmath}
\usepackage{palatino}
\usepackage{bigints}
\usepackage{mathpazo}
\usepackage{soul}
\usepackage{graphicx,amssymb,amsmath,amsthm,amsfonts,amscd, mathrsfs,epsfig,epsf}
\usepackage{textcomp}
\usepackage{float}
\usepackage{booktabs}
\usepackage{dcolumn}
\usepackage{booktabs}
\usepackage{multirow}
\usepackage{hyperref}
\hypersetup{colorlinks,citecolor=blue}
\usepackage{amsmath}
\usepackage[mathcal]{euscript}
\usepackage{xcolor}
\usepackage{orcidlink}
\usepackage{commath}
\usepackage{subcaption}
\def \nn  {\nonumber}
\def\be{\begin{eqnarray}}
\def\ee{\end{eqnarray}}

\def\jnl@style{\it}
\def\aaref@jnl#1{{\jnl@style#1}}

\def\aaref@jnl#1{{\jnl@style#1}}

\def\aj{\aaref@jnl{AJ}}                   
\def\apj{\aaref@jnl{ApJ}}                 
\def\apjl{\aaref@jnl{ApJ}}                
\def\apjs{\aaref@jnl{ApJS}}               
\def\apss{\aaref@jnl{Ap\&SS}}             
\def\aap{\aaref@jnl{A\&A}}                
\def\aapr{\aaref@jnl{A\&A~Rev.}}          
\def\aaps{\aaref@jnl{A\&AS}}              
\def\mnras{\aaref@jnl{Mon.~Not.~Roy.~Astron.~Soc.}}             
\def\prd{\aaref@jnl{Phys.~Rev.~D}}        
\def\prc{\aaref@jnl{Phys.~Rev.~C}}  
\def\prl{\aaref@jnl{Phys.~Rev.~Lett.}}    
\def\qjras{\aaref@jnl{QJRAS}}             
\def\skytel{\aaref@jnl{S\&T}}             
\def\ssr{\aaref@jnl{Space~Sci.~Rev.}}     
\def\zap{\aaref@jnl{ZAp}}                 
\def\nat{\aaref@jnl{Nature}}              
\def\aplett{\aaref@jnl{Astrophys.~Lett.}} 
\def\apspr{\aaref@jnl{Astrophys.~Space~Phys.~Res.}} 
\def\physrep{\aaref@jnl{Phys.~Rep.}}      
\def\physscr{\aaref@jnl{Phys.~Scr}}       
\def\commat{\aaref@jnl{Comm.~Math.~Phys.}}              
\def\science{\aaref@jnl{Science}}               
\def\cqg{\aaref@jnl{Classical Quant.~Grav.}}            
\def\jpcs{\aaref@jnl{JPCS}}                                     
\def\ijmpd{\aaref@jnl{Int.~J.~Mod.~Phys.~D}}                    
\def\grg{\aaref@jnl{Gen.~Relat.~Gravit.}}               
\def\rpp{\aaref@jnl{Rep.~Prog.~Phys.}}          
\def\npa{\aaref@jnl{Nucl.~Phys.~A}}        
\def\lrr{\aaref@jnl{Living Rev.~Rel.}}                   
\def\jcap{\aaref@jnl{J.~Cosmology Astropart.~Phys.}}    
\def\rmp{\aaref@jnl{Rev.~Mod.~Phys.}}   
\def\epjc{\aaref@jnl{Eur.~Phys.~J.~C}}

\DeclareMathOperator{\arccosh}{arc cosh}

\allowdisplaybreaks[1]

\begin{document}

\color{black}       

\title{Accelerated black holes in (2+1) dimensions: Quasinormal modes and Stability}

\author{R. D. B. Fontana  \orcidlink{0000-0002-8835-7447}}
\email[Email: ]{rodrigo.dalbosco@ufrgs.br}
\affiliation{Universidade Federal do Rio Grande do Sul, Campus Tramandaí-RS Estrada Tramandaí-Osório, Tramandaí CEP 95590-000, RS, Brazil} 
\affiliation{Departamento de Matematica da Universidade de Aveiro and Centre for Research and Development 
in Mathematics and Applications (CIDMA), Campus de Santiago, 3810-183 Aveiro, Portugal}
\author{Angel Rincon \orcidlink{0000-0001-8069-9162}}
\email[Email: ]{angel.rincon@ua.es}
\affiliation{Departamento de Física Aplicada, Universidad de Alicante, Campus de San Vicente del Raspeig, E-03690 Alicante, Spain}


\begin{abstract}
We investigate the propagation of a scalar field in a $(2+1)$-dimensional accelerated black hole, recently revisited in \cite{Arenas_Henriquez_2022}. We briefly describe the minimally-coupled configuration as rendering a trivial scalar perturbation with a rescale of the field mass. On the contrary, the free scalar field propagation presents an intricate dynamic, whose equation may be reduced through the use of Israel junction conditions and a non-trivial ansatz. In this case, using two different methods we calculate the quasinormal modes of the solution also obtaining unstable field profiles delivered by the linear scalar perturbation to the background geometry. We scrutinize the parameter space of angular eigenvalue of the field and accelerations under which such instabilities happen.

\end{abstract}

\maketitle

\section{Introduction}

Up to now, General Relativity (GR hereafter) gives the most acceptable description of gravity ~\cite{Einstein:1915ca}, raising its representation to a special spot over every other theory of gravity. 
Such privilege is supported by a vast number of predictions albeit two of them, i) gravitational waves (GWs hereafter) and ii) black holes (BHs) deserve special attention. The reason for that is self-evident: both represent an undeniable confirmation of Einstein's general relativity.
One decade ago, the direct detection of GWs with the emission of signals from a binary BH merger and the subsequent ringdown of the single resulting BH~\cite{LIGOScientific:2016aoc} was achieved for the first time, ushering a new era of gravitational physics data. 
Additionally, the detection of X-rays from intensely heated material swirling around a unknown objects provided strong evidence for a central black hole such as Cygnus X-1 in the Milky Way.
Motivated by the recent advances and findings in black hole physics, the study of GWs and BHs is now more relevant than ever and such studies are useful in unraveling the mysteries of the cosmos at a fundamental level. Moreover, it is known that 
BH physics can also be used to gain insight into the unification of GR, quantum theory, and statistical mechanics~\cite{Hawking:1975vcx,Bekenstein:1973ur,Bardeen:1973gs,Hawking:1982dh}, thus opening a new way to combine classical gravity with quantum mechanics. 

Given the current evidence, it is clear that a four-dimensional spacetime represents the natural case of interest. 

Considering the simpler structure when compared to their higher dimensional counterparts, lower dimensional theories of curvature (v. g. BTZ and Jackiw theories) are an interesting theoretical proposal, worth of investigation. The relevance of gravity in three dimensions becomes obvious because 
i) the absence of propagating degrees of freedom makes the spacetime simpler than in (3 + 1) dimensions, and
ii) the deep connection with Chern-Simons theory \cite{Witten:1988hc,Achucarro:1986uwr,Witten:2007kt} makes such exploration a vibrant field of research. 
Three-dimensional gravity is also convenient because of its mathematical simplicity (i.e., because it is quite easy to do the calculations), and because its features strongly resemble those in four-dimensional gravity theory. 
For the above reasons, curvature theories in $(2+1)$ dimensions have become very relevant in the last decades, mainly motivated by the now seminal paper by Bañados and coworkers (a.k.a. BTZ black hole ({\it{Bañados-Teitelboim-Zanelli}}) \cite{Banados:1992wn}, which was the first to compute a black hole in $(2+1)$ dimensions with a cosmological constant. Of equal importance are the solutions of the Jackiw gravity ~\cite{Jackiw:1984je} in $(1+1)$ dimensions (for a comprehensive survey see~\cite{Mann:1991qp}).

From then on, the field became relevant being convenient for learning complicated aspects of physics in four and higher dimensions as e. g. the connections delivered in the AdS/CFT correspondence. In this sense, lower-dimensional gravity is conventionally considered to be the perfect arena for gaining insights into four (and higher) dimensional gravity, namely black holes. 

Examples of black holes in three dimensions are abundant in the literature, where several properties are also computed (see for instance \cite{Birmingham:2001dt,Holst:1999tc,Birmingham:1998pn,Lee:1998pd,Dasgupta:1998jg,Cadoni:2010ztg,Cataldo:2000we} and references therein). Even more, substantiating the relevance of the topic, textbooks of such theories are now available as, e. g. \cite{Jackiw:1984je,Garcia-Diaz:2017cpv}.
Some of the scenarios investigated are quantum black holes (or more precisely, solutions with quantum effects) where some aspects such as geodesics, the Sagnac effect, thermodynamics, among others are computed (see \cite{Emparan:2020znc,deOliveira:2018weu,Koch:2016uso,Rincon:2017ypd,Contreras:2019iwm,Rincon:2019zxk,Fathi:2019jid,Rincon:2020izv,Panah:2022cay,Panotopoulos:2022bky,Rincon:2022hpy} and references therein), and some recent summaries of the progress made up to now can be consulted in the excellent textbook \cite{Carlip:1998uc}.


GR predicts the existence of black holes (BHs) and gravitational waves (GWs). In particular, the stability of the outer spacetime of classical black holes for small perturbations and quasinormal modes (QNMs) has long been an important goal of the theory \cite{Regge:1957td,Zerilli:1970se,Zerilli:1970wzz,Zerilli:1974ai,Moncrief:1975sb,Teukolsky:1972my} (see also the seminal monograph \cite{Chandrasekhar:1985kt}).
More precisely, perturbations in black holes can be categorized into three different phases: i) the generation of radiation in response to initial conditions, ii) subsequent damped oscillations characterized by complex frequencies, and iii) a power-law decay of the fields. The complex frequencies characterizing phase (ii), denoted as $\omega \equiv  \omega_R - i \omega_I$, are determined by a few key black hole parameters, namely i) mass, ii) electric charge, and iii) rotation, collectively referred to as the Quasinormal Modes (QNMs) of black holes. It is noteworthy that the real part of these frequencies governs the oscillation period, expressed as $T = 2\pi/\omega_R$, while the imaginary part encodes the fluctuation decay at a timescale of $t_D = 1/\omega_I$.
%
Following the groundbreaking direct observation of gravitational waves by LIGO in black hole mergers \cite{LIGOScientific:2016aoc,LIGOScientific:2016sjg,LIGOScientific:2017bnn}, we now possess the most compelling evidence to date supporting the existence and merging of black holes and QNMs. This discovery has not only confirmed the reality of black holes but has also unveiled an entirely new perspective on our understanding of the universe.
In the late stages of black hole formation, perturbations exhibit quasinormal modes, producing gravitational radiation in a damped sinusoidal waveform. The quasinormal frequency spectrum provides unique details about the parameters of the black hole, facilitating the determination of its mass and angular momentum. The study of quasinormal modes has recently gained attention for its importance in understanding spinning black holes.
Certainly, quasinormal modes hold considerable relevance in four dimensions. Nonetheless, exploring lower-dimensional spacetimes proves valuable as well, in gaining insights into how specific modified theories affect QN frequencies and its connections to semiclassical aspects of gravity.
Several works on QNMs (and greybody factors) in 2+1 dimensions can be consulted in \cite{Cuadros-Melgar:2022lrf,DalBoscoFontana:2023syy,Fontana:2023dix,Fernando:2022wlm,Ovgun:2018gwt,Fernando:2008hb,Becar:2019hwk,Anabalon:2019zae,Gonzalez:2021vwp,Rincon:2018ktz,Rincon:2018sgd,Panotopoulos:2016wuu} (and references therein).

Accelerating black holes have a plethora of properties which distinguish them from other black holes. First, the spacetime of these types of black holes come from the C-metric \cite{Kinnersley:1970zw,Plebanski:1976gy,Griffiths:2005qp}. The force driving the acceleration comes from a conical deficit angle along a polar axis of this black hole. Also, the asymptotic behaviour of accelerating black holes (described by the C-metric) depends on several parameters that can lead to an accelerating horizon and make the asymptotic structure more complicated. For several examples of accelerating black holes, see \cite{Appels:2016uha,Astorino:2016ybm,Anabalon:2018ydc,Gregory:2019dtq,Zhang:2018hms,Destounis:2020pjk,Ball:2021xwt,Zhang:2020xub,EslamPanah:2019szt,EslamPanah:2023rqw} and references therein.


Considering the importance of (2+1) gravity in general, and the BTZ black hole in particular and the recent increasing interest in accelerated black holes, we aim in this work to study the scalar perturbation and corresponding quasinormal modes of an accelerated BTZ background.

The paper is structured as follows. After this short and compact introduction, in Sec.\eqref{sec2} we will discuss the main features of the $(2+1)$ accelerated BTZ black hole solution. 
Then, in Sec.\eqref{sec3}, we will introduce the background to be considered in the corresponding scalar perturbations. In the same section, we compute the quasinormal modes and frequencies due to a massless scalar probe field and discuss the effect of the acceleration on the stability. 
Our numerical results, as well as some figures and tables, are delivered in Sec.\eqref{sec4}. Finally, in Sec.\eqref{sec5} we discuss our results and possible perspectives for future work.

\section{Background: the (2+1) dimensions accelerated BHs} \label{sec2}

We begin by introducing the accelerated version of the BTZ black hole without rotation and charge as described in \cite{Astorino_2011, Xu_2012, Arenas_Henriquez_2022, Cisterna_2023} by
\be
\label{b1}
ds^2 = \Omega^{-2}\Big( -P(y)d\tau^2 + P^{-1}(y)dy^2 + Q^{-1}(x) dx^2 \Big), \hspace{0.5cm}
\ee
with 
%
\begin{align}
    \begin{split}
        \Omega &= a (x-y), 
        \\
        P(y) &= \frac{1}{a^2L^2}+1-y^2, 
        \\
        Q(x) &= x^2-1 
\label{b2}
    \end{split}
\end{align}
In such geometry, we can insert a strut \cite{Arenas_Henriquez_2022} at a particular $x=x_s$ with induced metric $ds^2_I= \Omega^{-2}(-Pd\tau^2 + P^{-1}dy^2$, that allows for the interpretation of $a$ as the acceleration parameter of the geometry. The normal vector to the strut (used to define boundary conditions for motion equations) is written as
\be
\label{nv}
\mathbf{n} = -\frac{\sqrt{Q_s}}{\Omega}dx.
\ee
When put into the Israel Junction conditions \cite{israel1}, the geometry presents a negative tension for the strut in the form
\be
\label{b3}
\mathfrak{T} = -\frac{a}{4\pi} \sqrt{Q_s}
\ee
with $Q_s = x_s^2-1$. In order to find an simple black hole geometry, connected with the original BTZ proposal, we consider another coordinate transformation defined as
\be
\label{b4}
x &=& \cosh (m \varphi ) \\
\label{b5}
r &=& -\frac{m}{ay}\\
\label{b6}
\tau &=& \frac{t}{ma}
\ee
in which $m$ represents the constant related to the position of the strut in the geometry established based the above transformation as
\be
\label{b6b}
m = \frac{\arccosh (x_s)}{\pi}
\ee
The new coordinates bring the metric to a more suitable configuration,
\be
\label{b7}
ds^2= \Omega^{-2} \Big( -Fdt^2 + F^{-1}dr^2 + r^2 d\varphi^2 \Big), 
\ee
with $F$ and $\Omega$ given by
\begin{align}
F &= \frac{r^2}{L^2} + m^2(r^2A^2-1), \\
\label{e2}
\Omega &= 1 + A r \cosh (m \varphi).
\end{align}
and $a=mA$. Such solution casts a regular accelerated spacetime with an event horizon $r_h$ at $F=0$, or $r_h=mL(1-A^2)^{1/2}$ pushed by a strut ($A>0$). It has a smooth transition to the pure BTZ spacetime whenever $A \rightarrow 0$ with $m^2$ interpreted as the mass of the solution. 

With the transformations (\ref{b4}-\ref{b6}), $\varphi$ represents an angular coordinate mirrored along the $x$-axis through the strut position: $x=x_s$ whenever $\varphi  = \pm \pi$ and $x=x_{min} \equiv 1$ with $\varphi = 0$. In the complete spacetime with the $x$ angular coordinate, the two coordinates patch are defined limited at their boundary by the presence of the strut. For this reason, specific (Israel) boundary conditions have to be considered for the field perturbations as we may further see.

We can obtain other types spacetimes from the primeval line element (\ref{b1}) by changing the character of the constants and coordinates. Supposing a rotation in the acceleration parameter of type $A \rightarrow -A$: the operation reverses the sign of (\ref{nv}) and (\ref{b3}) and - with the proper coordinate adjustments - modify the structure of the spacetime turning its defect into a domain wall. We can also portray the transformation $m \rightarrow im$ which brings the solution yet to another accelerated geometry of the BTZ class (viz. a particle without a horizon \cite{Arenas-Henriquez:2023vqp}). In this geometry however, the BTZ solution is not smoothly recovered whenever the acceleration is turned of, and for that reason, we will restrict our analysis to that of an accelerated spacetime with a strut.

We finish this section by quoting relevant papers in the characterization of the accelerated black hole we here study: some thermodynamical aspects of the spacetime were studied in \cite{tian2024thermodynamics}, although seminal features as the first law remain an open subject in the literature; semiclassical properties of the scalar field in the spacetime in \cite{Anber_2008} and the scalar field as part of the background in \cite{Cisterna_2023}. Further aspects as the Cotton classification in terms of a null-system can be elucidated through \cite{podolsky2023novel,papajcik2023algebraic} and a non-linear version of the charged accelerated black hole can be found in \cite{Eslam_Panah_2023}.

\section{Scalar Perturbations} \label{sec3}

We aim to study the scalar perturbation in an accelerated BTZ black hole through the scattering of a free/minimally coupled scalar field delivered via matter term,
\be
\label{e3}
\mathcal{S}_m = \int_{\cal M} d^3x \sqrt{-g}\left(\partial_\mu \Phi \partial^\mu \Phi -\varepsilon \Phi^2 \right).
\ee
The motion for such scalar field action is given by
\begin{align}
\nn
\Box \Phi -\varepsilon \Phi &= g_{\alpha\beta} \nabla^\alpha\nabla^\beta \Phi -\varepsilon \Phi 
\\ \label{e4}
&= \frac{1}{\sqrt{-g}}\partial_\mu \Big( \sqrt{-g}g^{\mu \nu} \partial_\nu \Phi \Big)  -\varepsilon \Phi =0, 
\end{align}
Here we analyze the scattering problem of the scalar field in the geometry (\ref{b1}), in two different scenarios: the free scalar field ($\varepsilon = 0$ and the minimally coupled  with $\varepsilon =  \mathcal{R}/8$. Following  the prescription of \cite{Wald_b} we consider a transformation of $\Phi$ related to its non-minimally coupled configuration \cite{fontana_2022} written as
\be
\label{e5}
\Phi \rightarrow \Omega^{1/2}\psi . 
\ee
With the above relation we start with the free Klein-Gordon field (we will address the minimally coupled configuration in the next section) casting it into the motion equation as
\be
\label{e6}
\frac{1}{\sqrt{-\mathfrak{g}}}\partial_\mu \Big( \mathfrak{g}^{\mu \nu} \sqrt{-\mathfrak{g}}\partial_\nu \Big) \psi +\left( \frac{\mathcal{R}}{8\Omega^2}  - \frac{\mathcal{F}}{8} \right) \psi =0,
\ee
in which $\mathfrak{g}= \text{diag} (-F,F^{-1}, r^2)$, and 
\be
\label{e7}
\mathcal{R} = -\frac{6}{L^2}
\ee
is the Ricci scalar of the geometry and $\mathcal{F} = \mathcal{R} - 6m^2A^2$ the Ricci scalar of $\mathfrak{g}$. In order to put equation (\ref{e6}) into a more suitable form, we perform the usual field transformation $\psi \rightarrow \zeta/\sqrt{r}$ and implement the tortoise coordinate, $\partial_{r_*} = F\partial_r$ into its derivative operators, obtaining
\be
\nn
\Omega^2 \left( \frac{\partial^2}{\partial r_*^2}-\frac{\partial^2}{\partial t^2}\right)\zeta + \Omega^2 F \left( \frac{F}{4r^2}- \frac{\partial_r F}{2r} \right)\zeta \\
\label{e8}
+\Omega^2 F \left(\frac{\partial_{\varphi \varphi }}{r^2} - \frac{\mathcal{F}}{8} \right) \zeta + \frac{\mathcal{R}F}{8} \zeta =0. 
\ee
The above relation (\ref{e8}) reduces to the usual pure BTZ equation \cite{Cardoso_2001} when $A=0$ ($\Omega=1$).

For the purpose of integrating (\ref{e8}) we can expand the conformal factor in a vectorial basis $X_k= e^{k m \varphi}$,
\be
\label{e9}
\Omega^2= \sum_{k=-2}^{2} \alpha_k X_k 
\ee
with $\alpha_{-k} \equiv \alpha_k = \left( \frac{Ar}{k}\right)^k$ for $k > 0$ and $\alpha_{0}=1+2\alpha_{2}$. 

Now, in order to match the angular part of (\ref{e8}) with the above, we must take the Klein-Gordon field in a similar basis, with the expansion,
\be
\label{e10}
\zeta = \Psi \sum_{n_r} \Theta_n, 
\ee
in which $\Theta_n $ stands for the angular part of $\zeta$ and $\Psi$ the radial-temporal (or double-null) dependence.
We write $\Theta_n $ in terms of imaginary numbers $n= n_r + in_i$ spanned by the functions $X$ as $\Theta_n = X_{n_r + in_i}$. In the non-accelerated limit we may take $n_r=0$ and recover the usual field decomposition \cite{Cardoso_2001}. 

In order to rewrite Eq. \eqref{e8} conveniently, we can introduce new operators $\hat{\mathcal{O}}_i$ defined as 
\begin{align}
    \hat{\mathcal{O}}_1 &= \left( \frac{\partial^2}{\partial r_*^2}-\frac{\partial^2}{\partial t^2}\right)  + F \left( \frac{F}{4r^2}- \frac{\partial_r F}{2r} - \frac{\mathcal{F}}{8} \right), \\
\label{e11}
\hat{\mathcal{O}}_2 &=F \frac{\partial_{\varphi \varphi }}{r^2}, \\
\hat{\mathcal{O}}_3 &= \frac{\mathcal{R}F}{8},
\end{align}

such that equation (\ref{e8}) in terms of $\hat{\mathcal{O}}_i$ is pictured as
\be
\Omega^2 \Big( \hat{\mathcal{O}}_1 + \hat{\mathcal{O}}_2 \Big)\zeta + \hat{\mathcal{O}}_3 \zeta = 0. 
\ee
In the above relation, $\hat{\mathcal{O}}_1 $ and $\hat{\mathcal{O}}_2$ act in the angular part of the field allowing Eq. (\ref{e8}) to be separated in the angular basis of vectors $X_k$.
 
 We analyze the equation for every operator, considering that the sum is not limited (v. g., $m n_r \in \mathbb{Z}$, prescribing a field decomposition). In the end the equation is rearranged in each pair $\pm n_r$ that respects the angular boundary condition (\ref{e10b}). Starting with $\hat{\mathcal{O}}_1$, we have
\be
\label{e12}
\Omega^2 \hat{\mathcal{O}}_1 \zeta =  \hat{\mathcal{O}}_1 \Psi \Big( \sum_{k=-2}^{2} \alpha_k X_k \sum_{n_r} X_{n_r} \Big)  X_{in_i} 
\ee
Since the sum in $n_r$ is taken in every integer value $m n_r$, each one of the five terms $X_k \sum_{n_r} X_{n_r}$ can be reordered as a simple sum $\sum_{n_r} X_{n_r}$. Then
\be
\label{e13}
\Omega^2 \hat{\mathcal{O}}_1 \zeta =  (\alpha_0 +2\alpha_1 +2\alpha_2 ) \hat{\mathcal{O}}_1 \zeta .
\ee
Similarly, for $\hat{\mathcal{O}}_2$ we can write
\be
\label{e14}
\Omega^2 \hat{\mathcal{O}}_2 \zeta = \frac{F}{r^2}\Psi \sum_{k=-2}^{2} \alpha_k X_k \sum_{n_r} n^2 m^2X_n
\ee
and 
\begin{align}
\begin{split}
&X_{in_i} \sum_{n_r}\sum_{k=-2}^{2} \alpha_k X_k (n_r+in_i)^2  m^2X_{n_r} = \\
&\sum_{n_r} m^2X_n \Big(  \alpha_0 (n_r + in_i)^2 +\hspace{0.9cm} \\
&\alpha_1 (n_r + in_i - m^{-1})^2 +  \alpha_{-1} (n_r + in_i+m^{-1})^2 + 
\\
&\alpha_2 (n_r +in_i-2m^{-1})^2 + \alpha_{-2} (n_r+ in_i+2m^{-1})^2 \Big). 
\label{e15}
\end{split}
\end{align}
Since $\alpha_k=\alpha_{-k}$, we can write every term in the rhs of the above equation in terms of a new operator $\tau$ to appear in the scalar potential (viz. (\ref{e18})),
\be
\label{e15b}
\tau = (\tau_++\tau_-)/2, \hspace{1.0cm} \tau_\pm \equiv \sum_{j=0}^{2}\alpha_j N_j^{(\pm )} 
\ee
in which
\be
\label{e16}
N_j^{(\pm )} = \frac{2}{(2-j)!}\Big(\frac{j^2}{m^2}+(\pm n_r+in_i)^2\Big).
\ee
We still can simplify (\ref{e15b}) considering the cancellation of the imaginary term and rewrite
\be
\nn
\tau =  \sum_{j=0}^{2}\alpha_j  \frac{2}{(2-j)!}\Big(\frac{j^2}{m^2}+ n_r^2 - n_i^2\Big). 
\label{e16b}
\ee
In such case, the real and imaginary parts of the angular dependence act in opposite direction in the frequencies and the eigenvalue to be considered is the difference between both. 
Since both $n_r$ and $n_i$ are quantized by the continuity rules, we may rewrite the angular eigenvalue in a different set, $n_r^2 - n_i^2 \equiv \gamma$ emphasizing the difference between both integer numbers, to obtain

\begin{align}
\tau \equiv \sum_{j=0}^{2}\alpha_j  \frac{2}{(2-j)!}\Big(\frac{j^2}{m^2}+ \gamma \Big).
\end{align}
We discuss the allowed values for $\gamma$ in the next subsection. 

The last operator, $\hat{\mathcal{O}}_3$ acts trivially on $\zeta$ not modifying the angular part of it. After considering all three operators, we summarize the Klein-Gordon equation in the usual form,
\be
\label{e17}
 \left( \frac{\partial^2}{\partial t^2}-\frac{\partial^2}{\partial r_*^2} + \mathcal{V} (r) \right)\Psi = 0 ,
\ee
with $\mathcal{V}(r)$ acting as the effective potential,
\be
\label{e18}
\mathcal{V}(r) \equiv F\left( \frac{F}{4r^2} - \frac{\partial_r F}{2r}-\frac{\mathcal{F}}{8} + \frac{\frac{\mathcal{R}}{8}+\frac{m^2\tau}{r^2}}{\alpha_0+2\alpha_1+2\alpha_2}\right).
\ee
We emphasize the need of the two real eigenvalues summed in Eq. (\ref{e15b}) seen in the above potential: we can only satisfy the boundary condition (\ref{e10b}) once in each equation (\ref{e17}) two modes are concomitantly considered ($n_r$ and $-n_r$). 

In the region of integration of the scalar equation, $[r_h, \infty )$, (\ref{e18}) can be entirely positive, partially positive or entirely negative. In table \ref{ta1} we analyse the potential according to to three important points: its signal ($s_V$) in both asymptotic regions and the number of inflections.

\begin{table*}
\centering
\caption{Effective potential behavior for the range of accelerations of the black hole with $r_h=L=1$ and $\delta=0$.}
\begin{tabular}{cccc}
\hline
 Region &   $s_V$ $(r \rightarrow r_h^{(+)})$ & $s_V$ $(r \rightarrow \infty )$ & Number of inflections \\
	\hline \hline
$R_1$ ($A<0.5$) & $+$ & $+$ & 0\\
$R_2$ ($0.5<A<0.706$) & $-$ & $+$ & 1\\
$R_3$ ($0.706<A<0.781$) & $-$ & $-$ & 1\\
$R_4$ ($0.781<A<0.867$) & $+$ & $-$ & 2\\
$R_5$ ($A>0.867$) & $+$ & $+$ & 1 \\
    \hline  
    \end{tabular}
  \label{ta1}
\end{table*}

\begin{figure*}[t]
\begin{center}
\includegraphics[width=0.3\textwidth]{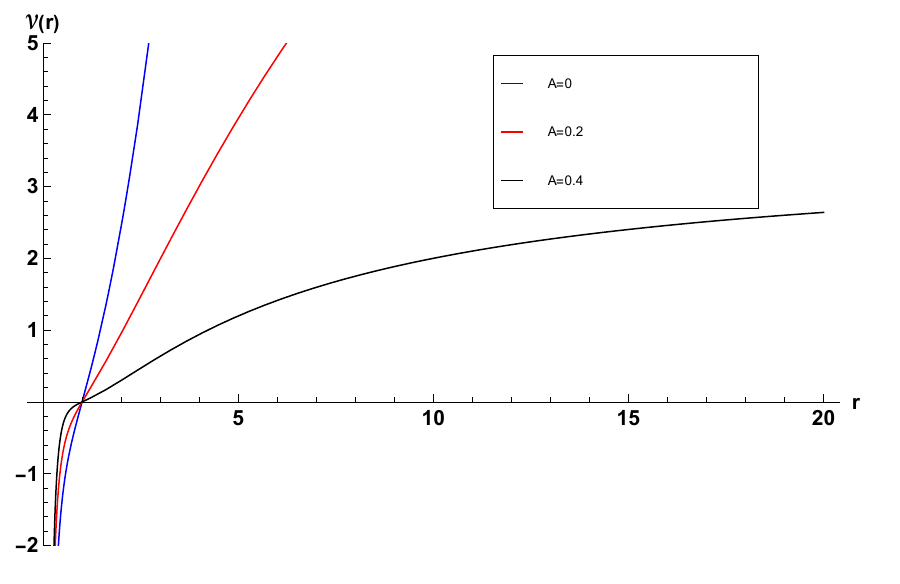}
\includegraphics[width=0.3\textwidth]{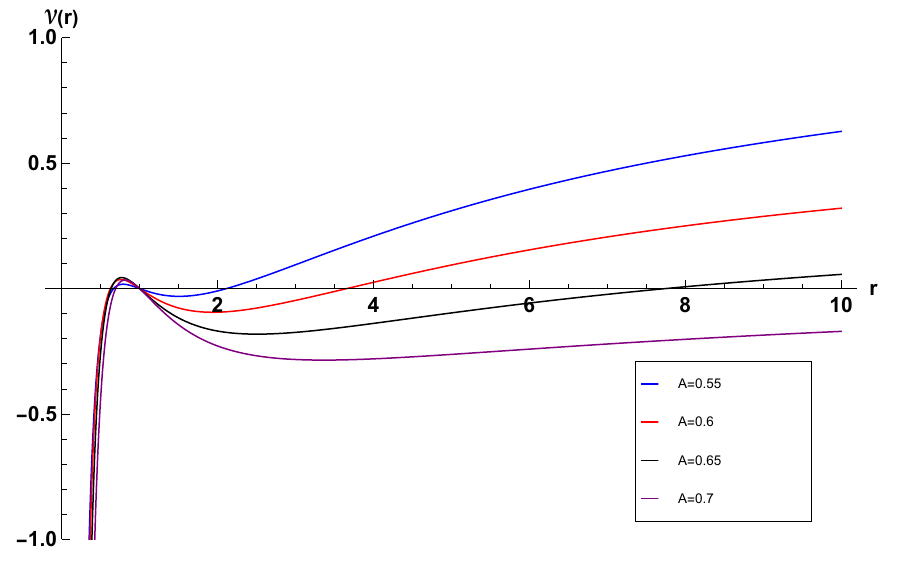}
\includegraphics[width=0.3\textwidth]{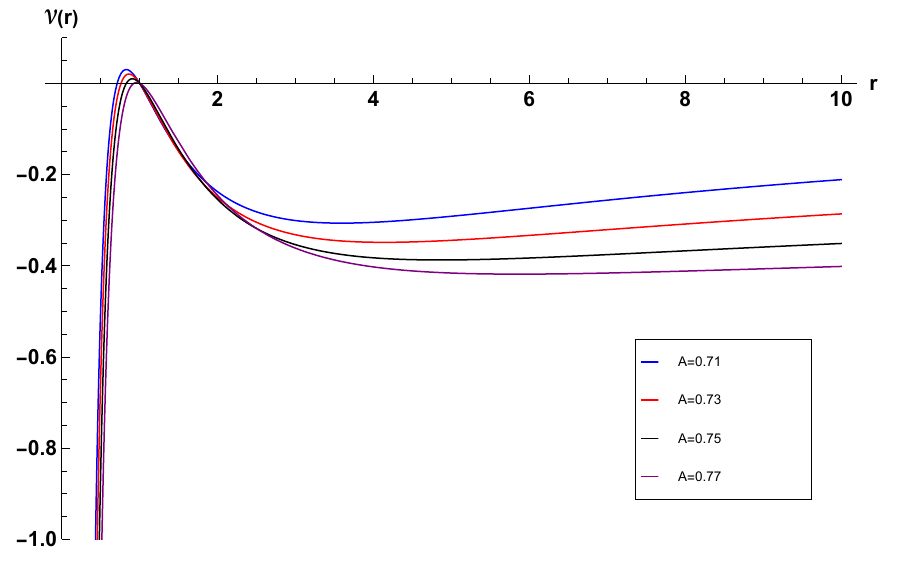}
\includegraphics[width=0.3\textwidth]{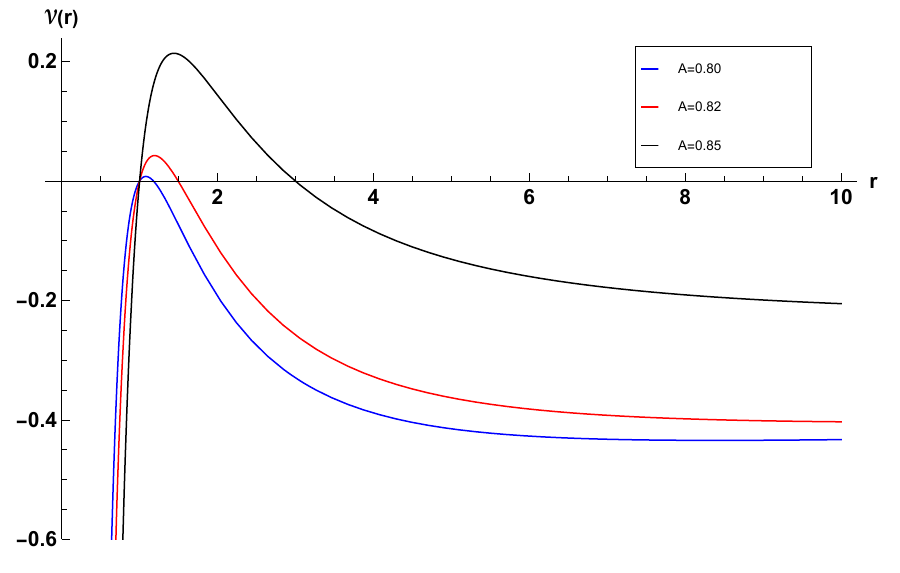}
\includegraphics[width=0.3\textwidth]{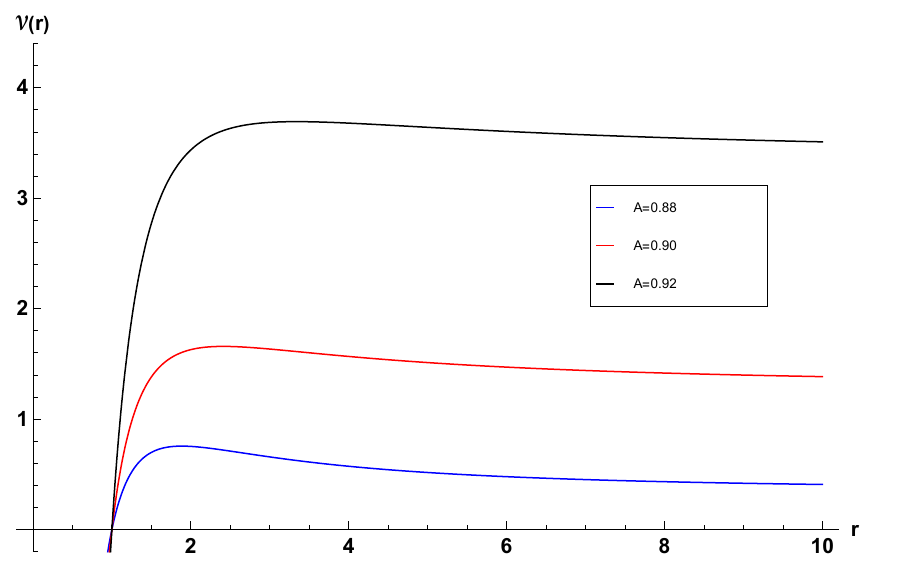}
\end{center}
\caption{The scalar field effective potentials for different geometric acceleration ($r_h=L=1$) with angular momentum $\delta=0$.}
\label{p1}
\end{figure*}

The rich structure of $\mathcal{V} (r)$ presented in \ref{ta1} demonstrates the non-trivial behavior of the propagation of the scalar field in such geometry. By instance, the emergence of an entirely negative potential for a range of accelerations ($R_3$) is expected to generate unstable profiles (see e. g.\cite{Horowitz_2000, Zhu_2014, Konoplya_2014, Fontana:2023dix, Fontana_2019} ). We may see in the next section that, that is not the case here. 
We represent all the qualitative cases of table \ref{ta1} in the plot \ref{p1}. 

Despite such rich and oddly shaped potentials, for entirely negative potentials unstable evolutions can not be taken for granted as wee will verify. We assemble the results regarding the dominant frequency and stability analysis in the next section. 

\subsection{Boundary Conditions} \label{sec4}

The scalar wave equation (\ref{e17}) allows us the investigation of the dynamical aspects of the spacetime considering the field at linear order. 

The evolution profile can be studied under specific boundary conditions: in the $r_*\rightarrow -\infty$ frontier ($r=r_h$), the field behavior is that of an ingoing plane wave, incident to the event horizon. That is the only boundary condition physically accepted in such classical scattering problems with motion equation similar to (\ref{e17}), whenever $\mathcal{V}(r)\Big|_{r_h} \rightarrow 0$, as long as no information can emerge from the inner region of the black hole. 

In spacetimes with AdS asymptotic structure as the BTZ geometries we are treating, different border terms are possible in such frontier. In \cite{Dappiaggi_2018} Robin boundary conditions are chosen, namely a zero energy flux in the AdS border. Other possible set of boundary conditions respecting different wave equations of AdS spacetimes is presented in \cite{Chen_2020}: plane waves in the AdS infinite modulated by a frequency $\sqrt{\omega^2 -\mathcal{V}_\infty}$, what only applies in cases where $\mathcal{V}_\infty$ is bounded. In the present spacetime we study, the condition of a plane wave in AdS infinite is not physical as long as $\mathcal{V}(r)$ is not smooth in that region when $A\rightarrow 0$. Also, as the Robin conditions \cite{Dappiaggi_2018} represents a general class of the usual $\Psi_{r_*\rightarrow 0} = 0$, this is the condition we use in the present work for the AdS border.

The physical boundary conditions for $\Theta_n$ require that $n_r m \in \mathbf{Z}$ as well as $n_i m \in \mathbf{Z}$, which is also imposed in the purely BTZ-like geometries ($n_r=0$). We still need the continuity of $\Theta_n$ along the limiting values of the $\varphi$ axis,
\be
\label{e10b}
\Theta_n(-\pi ) = \Theta_n (\pi )
\ee
In equation (\ref{e16b}) such condition narrows down the angular eigenvalue $\gamma$ ($\equiv \delta /m^2$) to another quantization rule, 
\be
\label{e10c}
 \delta \in \mathbf{Z_s}
\ee
in which $\mathbf{Z_s}$ is a subgroup of the integer numbers that excludes those multiples of $4\mathbf{Z}+2$, namely
\be
\label{e10d}
 \delta = \pm \{ 0, 
1,3,4,5,7,8,9,11,12,13, \cdots\}. \hspace{1.5cm}
\ee
The derivative of the field on the other hand is not  continuous in $\varphi$ (as the metric components are discontinuous as a consequence of the presence of a wall/strut). 

In such case, the Israel junction conditions \cite{israel1, Cisterna_2023, Poisson_2004} must be considered put into the form
\be
\label{eisr}
n^\mu [\partial_\mu \zeta]= n^\mu \partial_\mu \zeta^{+} - n^\mu \partial_\mu \zeta^{-} =  0.
\ee
where we recall that $n$ is the normal vector to the strut (\ref{nv}), located at $x=x_s$ and the $\pm$ signals refer to the terminal point of each coordinate patch, $\varphi^{\pm}=\pm \pi$. We emphasize the limits of $\varphi$ as mirroring coordinates to $x \in [1, x_s)$.

The field Ansatz (\ref{e10}) can not satisfy the boundary condition (\ref{eisr}). However, once we duplicate (\ref{e10}) considering a secondary wave with negative $n_i$ or
\be
\label{fa2}
\zeta = \Psi \Big(X_{in_i}+X_{-in_i}\Big)\sum_{n_r} X_{n_r} = 2\Psi \cos (n_i m \varphi) \sum_{n_r} X_{n_r} \hspace{0.8cm}
\ee
equation (\ref{eisr}) is fulfilled. Interestingly enough, the wave equation remains unaffected as in (\ref{e17}-\ref{e18}) when we consider the Ansatz (\ref{fa2}). The reason is that, summing another equation substituting $n_i \rightarrow -n_i$ does not change the term $\tau$ in the potential. 

\subsection{Numerics}
%
The integration of (\ref{e17}) is performed with an usual technique prescribed firstly in \cite{Gundlach_1994} by Gundlach, Price and Pullin. 
Firstly, we rewrite \eqref{e17} in double-null coordinates (usually defined, $dv=dr_*+ dt$ and $du=dt-dr_*$) obtaining
\begin{align}
 \left(
 4\frac{\partial^2}{\partial u \partial v} + \mathcal{V}(u,v)
 \right) \Psi(u,v) &= 0.
\end{align}
Using such coordinates, we can discretize the wave equation as \cite{Konoplya:2011qq}
\begin{align}
\Psi_N &= \Psi_E + \Psi_W - \Psi_S  - 
\frac{h^2}{8} \mathcal{V}_S \Bigl(\Psi_W + \Psi_E\Bigl), 
\end{align}
Such expression is quite efficient for asymptotically flat or de-Sitter black holes, but its convergence is relatively slow for asymptotically AdS black holes \cite{Konoplya:2011qq}. To circumvent this issue, an alternative integration scheme is 
\be
\nn
\Psi_N = \Big(1+\frac{h^2}{16}\mathcal{V}_S\Big)^{-1}\Big(\Psi_E + \Psi_W - \Psi_S \hspace{1.2cm} \\- 
\label{e21}
\frac{h^2}{16}(\mathcal{V}_S\Psi_S+\mathcal{V}_E\Psi_E+\mathcal{V}_W\Psi_W)\Big),
\ee
Afterwards we apply in (\ref{e21}) generic Cauchy data,
\be
\label{e21b}
\Psi_{r_* \rightarrow 0} = 0, \hspace{1.5cm} \Psi (u_0, v) = \text{gaussian} (v)
\ee
that allows for the field evolution and acquisition of the field profile after the above initial burst (\ref{e21b}) is imposed. With the signal at hand we finally can apply a spectroscopic technique as the Prony method \cite{Konoplya:2011qq}. It consists of filtering the signal with a specific number of {\it overtones}, $\nu$
\be
\label{e22}
\Psi = \sum_{j=1}^{\nu} C_j e^{-i \omega_j t}
\ee
to which the fundamental is the most expressive, (most influencing in the signal). The filtering is done 
dividing a particular time interval in the field evolution in a large number of steps and inverting (\ref{e22}) for a specific $\nu$. A thorough description can be found e. g. in \cite{Konoplya:2011qq}.
%
%
%
%

As a double check to our results, we used a secondary method to probe our results of the characteristic integration scheme, the Frobenius expansion as developed by Howoritz and Hubeny \cite{Horowitz_2000}. In the appendix \ref{ap1} we elaborate some of the equations needed to perfom the acquisition of the quasinormal frequencies.

\section{Results}

\subsection{Minimally coupled configuration}

The minimally coupled motion equation (\ref{e4}) of the scalar field in an accelerated BTZ black hole  (\ref{b1}) can be extremely simplified with the transformation (\ref{e5}). We can write
\be
\label{e24}
\Box_g \Phi -\frac{\mathcal{R}}{8} \Phi = \Box_{\tilde{g}} \Psi - \frac{\mathcal{F}}{8} \Psi = 0
\ee
with the conformal metric $\tilde{g}$ given by $d\tilde{s}^2 = \Omega^2 ds^2$, or $diag(-F,F^{-1},r^2)$. This system represents the same as the simplest BTZ solution with the cosmological term rescaled as $L^{-2} \rightarrow L^{-2} + m^2A^2$. In this case, we obtain two groups of solutions, the first one possessing the scaling proposed in \cite{bir1} when Dirichlet boundary conditions are considered,
\be
\label{e25}
\omega_{(D)} = \pm \frac{n_i}{L} - i\frac{r_h}{L^2}\Big(2\nu +1 + \sqrt{1 + \mu_{eff}^2}\Big) 
\ee
in which we emphasize the role played by the term 
\be
\label{e26}
\mu_{eff}^2 \equiv \mathcal{F}/8 = -\frac{3}{4}\left( 1 + m^2L^2A^2 \right)
\ee
as an effective mass of the scalar field.
The second one is that produced by Neumann boundary conditions representing in our case stable solutions \cite{Dappiaggi_2018},
\be
\label{e27}
\omega_{(N)} = \pm \frac{n_i}{L} - i\frac{r_h}{L^2}\Big(2\nu +1 - \sqrt{1 +  \mu_{eff}^2}\Big) 
\ee
Interestingly enough, for black holes with $m^2L^2A^2>1/3$, the mass term acts in the real part of the frequency, breaking the scaling between $\omega_R$ and the angular momentum of the field as seen in the pure BTZ solution \cite{Cardoso_2001}. We also notice the extra term in the frequencies as the distance of an accelerated version of the BTZ and the non-accelerated spacetime, since $m^2L^2A^2=m^2L^2 - r_h^2$.

\subsection{Free scalar field}

\begin{figure*}[t]
\begin{center}
\includegraphics[width=0.4\textwidth]{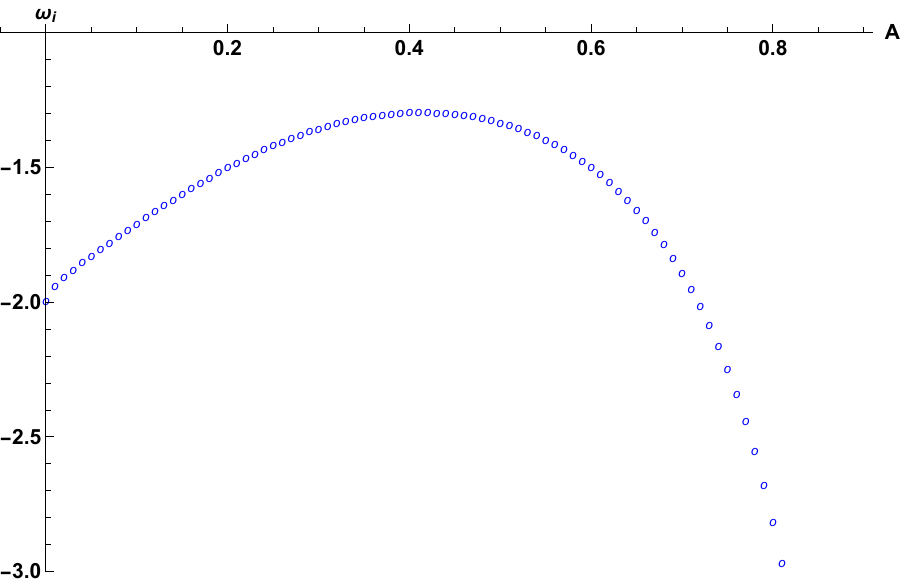}
\includegraphics[width=0.4\textwidth]{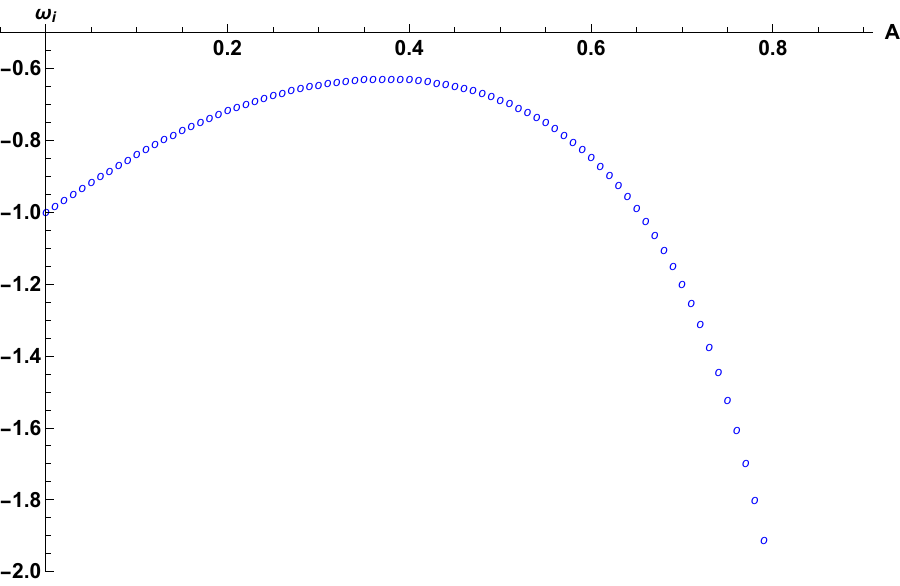}
\includegraphics[width=0.4\textwidth]{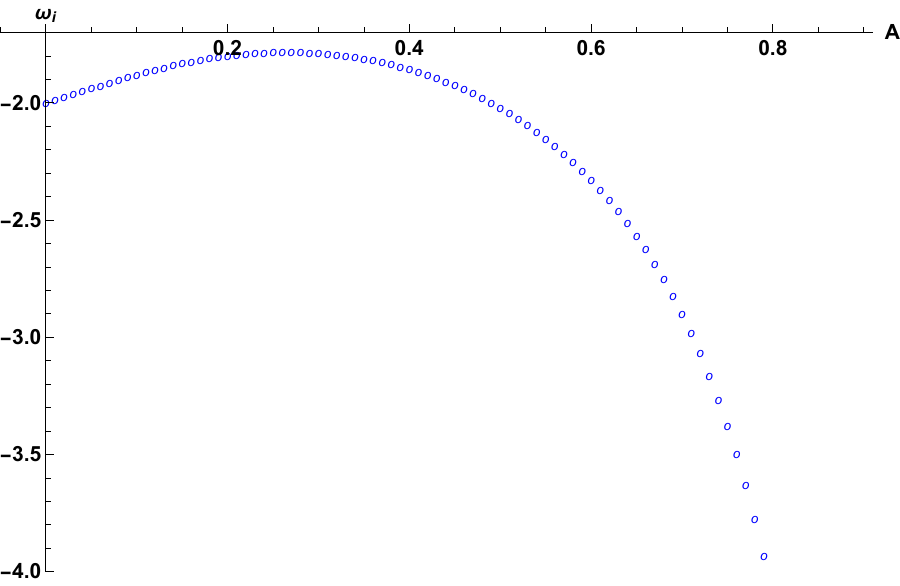}
\includegraphics[width=0.4\textwidth]{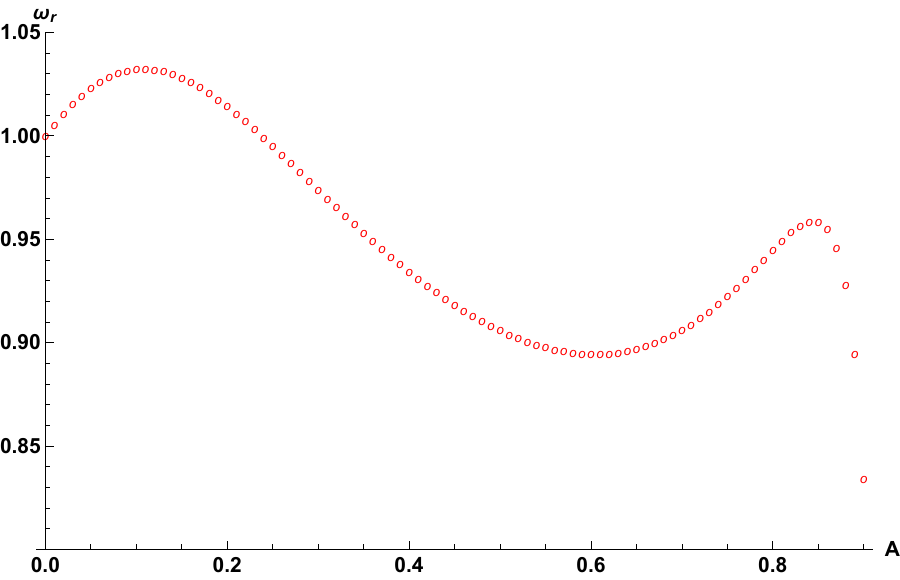}
\end{center}
\caption{Quasinormal modes with $\delta =0$ (top-left panel), $\delta =1$ (top-right panel) and $\delta =-1$ (bottom panels). The geometry parameters are $r_h=L=1$.}
\label{fd1}
\end{figure*}

The fundamental quasinormal modes of the free scalar field (\ref{e17}) can be put into two different sets, depending on the signal of the angular eigenvalue, $\delta$. In both cases, the field perturbations for small $|\delta|$ are stable. Its evolution occur as a tower of quasimodes after the initial burst. The resultant frequencies of such evolution are shown in figures \ref{fd1} for $|\delta| \leq 1$.

The results displayed at figures \ref{fd1} show interesting peculiarities of the scalar propagation in the accelerated spacetime. First of all the oscillatory nature of the modes is qualitatively unaffected by $A$, that is, oscillatory waves maintain their behavior ($\omega_r \neq 0$) in the accelerated spacetime, for every value of $A$, the same being true for purely imaginary evolutions. Second, the existence of positive decoupling constant $\delta$, behavior associated with the presence of the strut in the geometry, diminishes the fundamental frequencies when compared to that found  in \cite{Cardoso_2001}. 

The imaginary part of $\omega$ behaves qualitatively the same for small $|\delta|$, with increasing acceleration: $\omega_i$ achieves a maximum value depending on $\delta$ and monotonically decreases from this value on. Such behavior was found in other (2+1)D spacetimes, but associated with the change in $\omega_r$ (see e. g. \cite{Cuadros-Melgar:2022lrf} and references therein). 

An important aspect of the spacetime response to the field perturbation is the oscillation pattern (oscillatory or purely imaginary).  In our case such response is defined solely trough the angular constant. When $\delta < 0$ we observe decaying oscillatory profiles dominating the spectrum and whenever $\delta \geq 0$ ($\omega_r = 0$) we have a purely damped profile. 

In the accelerated BTZ black holes we observe instabilities for if $\delta >0$ is sufficiently large. In the pure BTZ limit, frequencies with $\delta >0$ were never reported as they result from non-usual boundary conditions (not described e. g. in \cite{Cardoso_2001}).

The presence of new modes unnoticed in the pure BTZ case whenever $\delta \leq 0$, can be summarized in the limit of very small acceleration ($mA\lesssim 0.01$) with the scale
\be
\label{e38}
\omega = \frac{\sqrt{-\delta}}{L} - 2\frac{r_h}{L}i,
\ee
which brings entire new towers of oscillations delimited by (\ref{e10d}), not bounded by the condition $\sqrt{-\delta} \in \mathbb{N} $ as in \cite{Cardoso_2001}. A table with the oscillatory frequencies for different $\delta <0$ is provided in \ref{tb2}. We notice the numerical results of the first line of \ref{tb2} that perfectly reproduce the scale of (\ref{e38}). 

\begin{table*}
  \centering
\caption{Quasinormal modes for negative $\delta$ with $r_h=L=1$.}
\scalebox{1.2}{\begin{tabular}{ccccc}
    \hline
 $A$ & $\delta = -1$ & $\delta = -3$ & $\delta = -4$ & $\delta = -5$ \\
	\hline \hline
0	&	0.9999-2.000i 	&	1.732-2.000i 	&	2.000-2.000i	&	2.236-2.000i	\\
0.1	&	1.032-1.879i	&	1.789-1.860i	&	2.063-1.854i	&	2.304-1.849i	\\
0.2	&	1.014-1.799i	&	1.793-1.779i	&	2.072-1.773i	&	2.317-1.768i	\\
0.3	&	0.9740-1.788i	&	1.789-1.777i	&	2.078-1.774i	&	2.331-1.771i	\\
0.4	&	0.9344-1.856i	&	1.804-1.857i	&	2.109-1.857i	&	2.375-1.857i	\\
0.5	&	0.9060-2.021i	&	1.853-2.033i	&	2.180-2.036i	&	2.464-2.038i	\\
0.6	&	0.8946-2.328i	&	1.952-2.350i	&	2.310-2.356i	&	2.620-2.360i	\\
0.7	&	0.9062-2.899i	&	2.130-2.930i	&	2.535-2.938i	&	2.885-2.944i	\\
0.8	&	0.9449-4.109i	&	2.461-4.148i	&	2.948-4.161i	&	3.367-4.170i	\\
    \end{tabular}}
  \label{tb2}
\end{table*}

The perturbations with $\delta >0$ bring the most important issue associated to the scalar field propagation in such spacetime: for $\delta \geq 3$ and a range of values of acceleration, instabilities are present. We list the quasinormal modes and exponential coefficients in table \ref{tb3}.

\begin{table*}
  \centering
\caption{Quasinormal modes for $\delta \geq 0$ with $r_h=L=1$.}
\scalebox{1.2}{\begin{tabular}{cccccc}
    \hline
 $A$ & $\delta = 0$ & $\delta = 1$ & $\delta = 3$ & $\delta = 4$ & $\delta = 5$ \\
	\hline \hline
0	&	-1.996i	&	-0.9993i	&	-0.2681i	&	$~0$	&	0.2358i 	\\
0.1	&	-1.708i	&	-0.8387i	&	-0.09255i	&	0.1807i	&	0.4212i 	\\
0.2	&	-1.500i	&	-0.7162i	&	0.02688i	&	0.3011i	&	0.5431i 	\\
0.3	&	-1.356i	&	-0.6444i	&	0.09475i	&	0.3709i	&	0.6154i 	\\
0.4	&	-1.296i	&	-0.6302i	&	0.1140i	&	0.3957i	&	0.6458i 	\\
0.5	&	-1.334i	&	-0.6863i	&	0.07773i	&	0.3703i	&	0.6310i	\\
0.6	&	-1.500i	&	-0.8460i	&	-0.04081i	&	0.2707i	&	0.5490i	\\
0.7	&	-1.890i	&	-1.199i	&	-0.3193i	&	0.02423i	&	0.3319i 	\\
0.8	&	-2.814i	&	-2.038i	&	-1.019i	&	-0.6172i	&	-0.2565i	\\
    \end{tabular}}
  \label{tb3}
\end{table*}

In this table, we see oscillations with negative imaginary parts representing quasinormal modes and others with positive coefficient portraying instabilities. The unstable profiles occur in different range of the space of parameters $\delta$ and $A$. Interestingly, for each $\delta >1$ there is a minimum ($A_i$) and a maximum ($A_s$) acceleration that triggers unstable evolution. We summarize in table \ref{tb4}.

\begin{table}[h]
  \centering
\caption{Critical values of acceleration for the scalar instability threshold.}
\begin{tabular}{ccc}
    \hline
\vspace{0.1cm}
 $\delta$ & $A_i$ & $A_s$ \\
3 & 0.18 & 0.57 \\
4 & 0  &  0.70 \\
5 & 0  & 0.76 \\
\hline
    \end{tabular}
  \label{tb4}
\end{table}

We must emphasize the convergence of the results in tables \ref{tb2}-\ref{tb4} with those obtained in the Frobenius expansion elaborated in the appendix of this work. For $\delta = 0$, the deviation in the fundamental modes does not surpass 0.4\% (except in the high acceleration regime) and in most of the other cases (with $\delta \neq 0)$ it is limited to 1\% to 2\%. The qualitative behavior of the modes is essentially the same as calculated with both methods, with the existence of instabilities being assured in both cases. 

The only exception for such good convergence in the frequencies between the results of both methods lies in the transient regions from stable to unstable evolutions, for high $\delta$. Such fact is actually expected as in the threshold regions, the convergence of the Frobenius method is highly affected for the series expansion considered with the nearly stationary configuration, $\omega \sim \omega_R - 0i$ (viz. $\omega_I \sim 0$).

It is worth mentioning however, that the critical values for the accelerations of table \ref{tb4} are confirmed in both methods within a deviation of 2\%, raising no doubts on the instabilities to linear perturbations on the accelerated spacetime in (2+1) dimensions. 

\section{Final remarks} \label{sec5}

In this work, we studied the scalar field perturbation in accelerated (2+1) dimensional BTZ geometries. The black hole possesses a topological defect (angular deficit) that results in a strut pushing the spacetime.  

Two different field configurations are analyzed, the non-minimally coupled and the free scalar field, with  different outcomes. 

The non-minimally coupled pattern brings the motion equation to the same as that of the free scalar field in non-accelerated BTZ spacetimes with an effective mass written in terms of the geometric acceleration as in Eq. (\ref{e26}).

The free scalar field representation unleashes a more subtle dynamic. In such regime, we must perform a non-usual field transformation (Ansatz) that leads to a decoupled motion equation (despite the acceleration of the spacetime). The non-trivial boundary conditions are governed by the Israel junction conditions in the patches delimited by the strut.

The decoupled equation was treated with the usual techniques available in the literature (characteristic integration and Frobenius method), with good convergence for the results (quasinormal modes and unstable frequencies) of both methods.

The quasinormal modes ($\delta \leq 0$) maintain the oscillatory pattern considering spacetimes with different accelerations, contrary e. g. to the charged BTZ spacetime \cite{DalBoscoFontana:2023syy}. The damping of the field achieves a minimum for intermediate accelerations growing from that point on for increasing $A$, achieving values considerate high when compared to other BTZ spacetimes.  

The results announce an unexpected behavior of the scalar field to linear order, its instability for high enough acceleration of angular eigenvalue. Instabilities triggered by a probe scalar field with non-trivial couplings scenarii \cite{Fontana_2019, Abdalla_2019} and/or non-usual global properties are well-represented in the literature from recent \cite{cardoso2024black} to ancient \cite{Gregory_1993} works. To this point it is unclear if the mechanism leading to the particular instability we present in this paper is a liberation mechanism as in the superradiant cases (supporting the evolution of scalar clouds as resolution process to the instability) or an elaborated channel perpetuated at high scale in the entire spacetime that modifies its background as a whole. The answer to such question must be investigated considering a fully non-linear evolution of the perturbations and is outside the scope of our work.

Further lines of investigation include: i) field perturbations with higher spin and the thermodynamics of scalar field in 2+1 dimensional accelerated geometries, or still the 
quantum-inspired scalar fields on a three-dimensional accelerated geometry.

\section*{Acknowledgments}
A. R. acknowledges financial support from the Generalitat Valenciana through PROMETEO PROJECT CIPROM/2022/13.
A. R. is funded by the María Zambrano contract ZAMBRANO 21-25 (Spain) (with funding from NextGenerationEU).

R. D. B. Fontana thanks the hospitality of DMAT - University of Aveiro/CIDMA  and the support of the Center for Research and Development in Mathematics and Applications (CIDMA) through the Portuguese Foundation for Science and Technology (FCT— Fundação para a Ciência e a Tecnologia), references https://doi.org/10.54499/UIDB/04106/2020 and https:// doi.org/10.54499/UIDP/04106/2020.

The authors also would like to thank to prof. Carlos Herdeiro for insightful suggestions and very fruitful discussions. We are also indebted for enlightening discussions with  Dr. Gabriel Arenas-Henriquez for what we are thankful.

\appendix

\section{Frobenius method}\label{ap1}

In order to apply the Frobenius method \cite{Horowitz_2000} in the wave equation (\ref{e17}) with potential (\ref{e18}), firstly we change the radial coordinate to a more suitable to the boundaries of the spreading problem,
\be
\label{ap1}
u = \frac{r_h}{r}
\ee
such that $u_h = 1$ and $u_\infty = 0$.  In this system the scalar motion equation turns to
\be
\nn
\frac{f^2u^4}{r_h^2} \frac{\partial^2 \Psi}{\partial u^2} + f^2u\left(\frac{fu}{r_h^2} -\frac{\Delta}{r_h} \right)\frac{\partial \Psi}{\partial u} + (\omega^2 - V)\Psi = 0 \\
\label{ap2}
\ee
in which $\Delta = (\partial_r F )_x$, $f=F_x$, $V = \mathcal{V}_x$ and $x$ represents the change of coordinate $r\rightarrow u^{-1}r_h$ in each function. The method consists in implementing the expansion
\be
\label{ap3}
\Psi =\sum_{n=0}^{N}(u-1)^{n +\alpha}
\ee
in the wave equation, solving the boundary condition near the event horizon implicitly, v. g. choosing the proper $\alpha$ for an ingoing plane wave. To leading order, equation (\ref{ap2}) produces
\be
\label{ap4}
\alpha =  \mp \frac{\omega }{\Delta} i.
\ee
The upper signal represents an ingoing wave while the bottom accounts for an outgoing that we may ignore. 

In a second step, we rewrite equation (\ref{ap2}) performing expansions for each function of the equation as
\be
\label{ap5}
s(u) \frac{\partial^2 \Psi}{\partial u^2} + \tau (u) \frac{\partial \Psi}{\partial u} + \Theta (u) \Psi = 0 
\ee
with
\begin{align}
\begin{split}
s (u) &= \frac{u^4 f^2}{r_h^2} = \sum_{n=0}s_n (u-1)^n \\
\tau (u) &= \frac{u^2 f}{r_h^2}(2uf-r_h\Delta) = \sum_{n=0}\tau_n (u-1)^n
\\
\Theta (u) &=  \sum_{n=0}\Theta_n (u-1)^n
\label{ap6}
\end{split}
\end{align}
and
\begin{align}
\begin{split}
\Theta = &\omega^2 + m^2\frac{u^2-1}{u^2} 
\Bigg( 
        - \frac{m^2u^2}{2r_h^2} -\frac{3u^2}{4L^2(u+Ar_h)^2)} 
\\
&+ \frac{4\delta u^2 + 8Ar_h u(1+\delta ) + 4A^2 r_h^2 (2+\delta )}{4r_h^2(u+Ar_h)^2}u^2 \Bigg) 
\label{ap7}
\end{split}
\end{align}
Finally, solving (\ref{ap5}) order by order, we obtain the recurrence relation for the coefficients $a_n$,
\be
\nn
a_n = -\frac{1}{(n^2+2n\alpha)s_2}\sum_{j=0}^{n-1}a_j\Big( \Theta_{n-j} + (\alpha + j )\tau_{n+1-j} +\\
\label{ap8}
(\alpha^2 + \alpha (2j-1) + j^2-j)s_{n+2-j} \Big)
\hspace{0.7cm}
\ee
The acquisition of the frequencies is done by constructing an algorithm that solves the imaginary polynomial equation derived from the quasinormal condition near infinity, $\Psi_\infty \rightarrow 0 $, $\Psi (u=0) = 0$ or
\be
\label{ap9}
\sum_{n=0}^{N}a_n (-1)^n =0
\ee
for a particular $N$. For such we used the well-known Muller`s procedure.
\newpage

\bibliographystyle{apsrev4-1}
\bibliography{Library}

\end{document}